\documentclass[manuscript]{emulateapj}

\slugcomment{Submitted to the Astrophysical Journal}
\shorttitle{FIR Observations of the Cas~A region}
\shortauthors{Sibthorpe et al.}
\usepackage{ae} 
\usepackage{aas_macros}
\usepackage[T1]{fontenc}
\usepackage[ansinew]{inputenc}
\usepackage{natbib}
\usepackage[FIGTOPCAP,nooneline]{subfigure}
\usepackage{multirow}
\usepackage{amsmath}
\usepackage{amssymb}
\usepackage{graphicx}
\usepackage{color}
\DeclareInputText{128}{\euro} 
\def\subsun{\mbox{$_{\odot}$}}
\def\arcsec{\hbox{$^{\prime\prime}$}}
\def\arcmin{$^{\prime}$}
\def\deg{\hbox{$^\circ$}}
\def\micron{$\mu$m}

\begin{document}

\title{AKARI and BLAST Observations of the Cassiopeia A Supernova Remnant and Surrounding Interstellar Medium}

\author{B.~Sibthorpe,\altaffilmark{1}
P.~A.~R.~Ade,\altaffilmark{2}
J.~J.~Bock,\altaffilmark{3,4}
E.~L.~Chapin,\altaffilmark{5}
M.~J.~Devlin,\altaffilmark{6}
S.~Dicker,\altaffilmark{6}
M.~Griffin,\altaffilmark{2}
J.~O.~Gundersen,\altaffilmark{7}
M.~Halpern,\altaffilmark{5}
P.~C.~Hargrave,\altaffilmark{2}
D.~H.~Hughes,\altaffilmark{8}
W.-S. Jeong,\altaffilmark{9}
H.~Kaneda\altaffilmark{10}
J.~Klein,\altaffilmark{6}
B.-C. Koo,\altaffilmark{11}
H.-G. Lee,\altaffilmark{12,11}
G.~Marsden,\altaffilmark{5}
P.~G.~Martin,\altaffilmark{13,14}
P.~Mauskopf,\altaffilmark{2}
D.-S. Moon,\altaffilmark{14}
C.~B.~Netterfield,\altaffilmark{14,15}
L.~Olmi,\altaffilmark{16,17}
E.~Pascale,\altaffilmark{2}
G.~Patanchon,\altaffilmark{18}
M.~Rex,\altaffilmark{6}
A.~Roy,\altaffilmark{14}
D.~Scott,\altaffilmark{5}
C.~Semisch,\altaffilmark{6}
M.~D.~P.~Truch,\altaffilmark{6}
C.~Tucker,\altaffilmark{2}
G.~S.~Tucker,\altaffilmark{19}
M.~P.~Viero,\altaffilmark{14}
D.~V.~Wiebe,\altaffilmark{5}}

\altaffiltext{1}{UK Astronomy Technology Centre, Royal Observatory Edinburgh, Blackford Hill, Edinburgh, EH9~3HJ, UK; {\url{bruce.sibthorpe@stfc.ac.uk}}}
\altaffiltext{2}{Department of Physics \& Astronomy, Cardiff University, 5 The Parade, Cardiff, CF24~3AA, UK}
\altaffiltext{3}{Jet Propulsion Laboratory, Pasadena, CA 91109-8099}
\altaffiltext{4}{Observational Cosmology, MS 59-33, California Institute of Technology, Pasadena, CA 91125}
\altaffiltext{5}{Department of Physics \& Astronomy, 6224 Agricultural Road, University of British Columbia, Vancouver, BC V6T~1Z1, Canada}
\altaffiltext{6}{Department of Physics and Astronomy, University of Pennsylvania, 209 South 33rd Street, Philadelphia, PA 19104}
\altaffiltext{7}{Department of Physics, University of Miami, 1320 Campo Sano Drive, Carol Gables, FL 33146}
\altaffiltext{8}{Instituto Nacional de Astrof\'{i}sica, \'{O}ptica y Electr\'{o}nica, Luis Enrique Erro 1, Tonantzintla, Puebla, 72840, Mexic}
\altaffiltext{9}{Korea Astronomy and Space Science Institute, 61-1, Hwaam-dong, Yuseong-gu, Daejeon 305-348, Korea}
\altaffiltext{10}{Department of Astrophysics, Nagoya University, Chikusa-ku, Nagoya 464-8602, Japan}
\altaffiltext{11}{Department of Physics and Astronomy, Seoul National University, Seoul 151-747, Korea}
\altaffiltext{12}{Department of Astronomy, Graduate School of Science, University of Tokyo, Bunkyo-ku, Tokyo 113-0003, Japan}
\altaffiltext{13}{Canadian Institute for Theoretical Astrophysics, University of Toronto, 60 St. George Street, Toronto, ON M5S~3H8, Canada}
\altaffiltext{14}{Department of Astronomy \& Astrophysics, University of Toronto, 50 St. George Street, Toronto, ON M5S~3H4, Canada}
\altaffiltext{15}{Department of Physics, University of Toronto, 60 St. George Street, Toronto, ON M5S~1A7, Canada}
\altaffiltext{16}{Istituto di Radioastronomia, Largo E. Fermi 5, I-50125, Firenze, Italy}
\altaffiltext{17}{University of Puerto Rico, Rio Piedras Campus, Physics Dept., Box 23343, UPR station, San Juan, Puerto Rico}
\altaffiltext{18}{Laboratoire APC, 10, rue Alice Domon et L{\'e}onie Duquet 75205 Paris, France}
\altaffiltext{19}{Department of Physics, Brown University, 182 Hope Street, Providence, RI 02912} 

\begin{abstract}
We use new large area far infrared maps ranging from 65-500\,$\mu$m obtained with the AKARI and the Balloon-borne Large Aperture Submillimeter Telescope (BLAST) missions to characterize the dust emission toward the Cassiopeia~A supernova remnant (SNR). Using the AKARI high resolution data we find a new ``\emph{tepid}'' dust grain population at a temperature of $\sim35$\,K and with an estimated mass of 0.06\,M$_{\odot}$.  This component is confined to the central area of the SNR and may represent newly-formed dust in the unshocked supernova ejecta.  While the mass of tepid dust that we measure is insufficient by itself to account for the dust observed at high redshift, it does constitute an additional dust population to contribute to those previously reported.  We fit our maps at 65, 90, 140, 250, 350, and 500\,$\mu$m to obtain maps of the column density and temperature of ``cold'' dust (near 16~K) distributed throughout the region. The large column density of cold dust associated with clouds seen in molecular emission extends continuously from the surrounding interstellar medium to project on the SNR, where the foreground component of the clouds is also detectable through optical, X-ray, and molecular extinction.  At the resolution available here, there is no morphological signature to isolate any cold dust associated only with the SNR from this confusing interstellar emission. Our fit also recovers the previously detected ``hot'' dust in the remnant, with characteristic temperature 100~K.
\end{abstract}


\section{Introduction}\label{The_CasA_SNR}
Determining the origin of cosmic dust is fundamental to our
understanding of many astronomical processes, including star formation
and galaxy evolution.
Galaxies and quasars at high redshift have been found to contain large
amounts of dust \citep[$\ge10^8$\,M\subsun;][]{Dunlop_1994,
  Archibald_dust_in_distant_quasars_2001, Isaak_2002, Priddey_2008},
at a time when the Universe was only about one tenth of its present
age.
The main source of dust \textit{injection} within our Galaxy is
thought to be the stellar winds of stars on the asymptotic giant
branch (AGB) of the Hertzsprung-Russell (HR) diagram
\citep{morgan_edmunds_2003}.
Stars at this early epoch would not have been able to reach the AGB
phase in the available time, and therefore cannot be the source of the
observed dust.
Heavy elements are produced in the explosions of supernovae (SNe) and,
for several years, models have predicted that considerable amounts of
fresh dust (0.1-1\,M\subsun) could also be produced
\citep{Kozasa_1991_SNR_dust_model, Woosley_1995_SNR_dust_model,
  Clayton_1999_SNR_dust_model, Todini_2001_SNR_dust_model}.
The life cycle of high mass stars ($>$8\,M\subsun), the progenitors of
type II SNe, is sufficiently short for SNe to occur within the
required timescales. As a result, SNe have been proposed as a possible
solution for the origin of the dust seen at high-redshift. However, in order for SNe to generate sufficient dust mass to fill this gap, each supernova would need to generate 0.4-1\,M\subsun~dust \citep{Dwek_2007}. This quantification does not account for dust grain destruction within the supernova, thereby making it a lower bound.

In seeking evidence regarding this hypothesis, the focus has been on
dust detectable in SNe and supernova remnants (SNR), as reviewed
briefly below.  However, it seems less well appreciated that injection is
only part of the story.  \citet{Draine_2003} reviews the often-ignored
arguments that, at least in our Galaxy, the interstellar dust is
continually processed on a timescale $3 \times 10^8$~yr and most of its
mass is formed in the interstellar medium. Nevertheless, injected dust
is critical at least as seeds for further evolution of the dust
population.

Studies of SNe find only trace amounts of dust in the hot ejecta, with typical masses of order 10$^{-4}$\,M$_{\odot}$ \citep{Dwek_1987,Lagage_1996,Arendt_1999,Ercolano_2007,Meikle_2007}.  Dust studies in SNR appear more promising, a prime example being Cassiopeia
A (Cas~A).
%
Cas~A is the remnant of a type IIb supernova event which occurred around AD 1680 \citep{Raymond_1984, Thorstensen_2001, Fesen_2006a, Krause_2008_TypeIIb}. The progenitor star is believed to have had a mass greater than 20\,M\subsun~\citep{Perez-Rendon_2002}, and the remnant is at a distance $D \sim 3.4$\,kpc \citep{Reed_1995}. Early observations of Cas~A made with IRAS \citep{IRAS_1984} and ISO \citep{ISO_1996} did not extend to longer wavelengths, and therefore detected only the ``hot'' ($\sim$100\,K) dust component, whose mass seemed insufficient to provide the levels of dust seen in high redshift galaxies. The low angular resolution also made the study of sub-structure in the remnant at intermediate wavelengths difficult.
The \textit{Spitzer Space Telescope} \citep{Spitzer_06} has been used
to study Cas~A \citep{Hines2004,Krause_2004_CasA,ennis2006}. Most
recently \cite{Rho_2008}, exploiting the angular and spectral
resolution achieved with the \textit{Spitzer} infrared spectrograph
\citep[IRS;][]{Houck_2004}, find between 0.020 -- 0.054\,M$_{\odot}$
of hot dust.  This is an order of magnitude greater than that
previously measured and they conclude that, within modeling
uncertainties for galaxy evolution, this could be sufficient to
explain at least the lower limit to the dust levels in high-redshift
galaxies presented by \cite{Isaak_2002}.

Much more controversial is the question of a ``cold'' dust component
in the SNR, because of the issue of contamination by line-of-sight
interstellar emission from dust which is also cold ($\sim$16\,K; \S~\ref{comp_sep}).
\cite{Dunne_2003_CasA}, using data from the Submillimetre Common User
Bolometer Array \citep[SCUBA;][]{Holland_SCUBA_99}, find evidence for
$2-4\,{\rm M\subsun}$ of dust in Cas~A at a temperature of
$\sim$18\,K, significantly more than the mass of hot dust.
Using the same methodology, \citet{Morgan_2003_Kepler} find $\sim$1\,M$_{\odot}$ of cold dust in Kepler's supernova remnant, a thousand times greater than previous measurements for this SNR. Both of these remnants are sufficiently young for the dust observed to be freshly formed in the remnant, rather than being material swept up from the ISM by the shock-wave \citep{Dickel_1988, Hughes_1999, Synch_Wright_1999}.
More recently, \cite{Dunne_2009} presented further evidence for cold
dust in Cas~A using SCUBA polarization data. These show dust emission
polarized in an orientation consistent with that of the magnetic field
deduced from the radio synchrotron emission, suggesting the detected
dust is in the SNR. They find a conservative lower limit for this dust
mass of 1\,M$_{\odot}$.  They also attribute apparent depolarization
at the brightest feature to dilution by line of sight interstellar
emission.

\cite{Dwek_2004} has argued that the mass estimates in
\cite{Dunne_2003_CasA} exceed the total anticipated mass ejection of
Cas~A and suggest that if the submillimeter observations are valid
they might instead imply the presence of a much smaller amount of dust
which is a much more efficient millimeter wavelength radiator, such as
iron needles.  The alignment of such needles could affect the polarization.

By correlating cold dust emission with molecular line absorption
against the SNR synchrotron emission, \cite{Krause_2004_CasA} argued
that the dust is, in fact, associated with a molecular cloud located
along the line of sight to the SNR.  Molecular emission has also been
studied in this direction (e.g., \citealt{liszt_lucas_1999}), showing
that the cloud(s) extend well beyond the SNR itself.
\citeauthor{Krause_2004_CasA} estimate the fresh dust yield within
Cas~A to be at least an order of magnitude lower than that found by
\cite{Dunne_2003_CasA}. Even so, this would still provide the
predominant dust mass in Cas~A, bolstering the possibility of
explaining the quantities of dust seen at high-redshift.

The emission from cold dust peaks in the far-infrared and
submillimeter and so neither SCUBA, on the long wavelength side, nor
Spitzer, on the short side, is ideal for isolating a cold dust
component.  Near the thermal peak, the best large scale maps covering
Cas~A and its environs are from the Multiband Imaging Photometer for
Spitzer \citep[MIPS;][]{MIPS_2004} at 160\,$\mu$m
\citep{Krause_2004_CasA} and from the ISO Serendipity Survey at
170\,$\mu$m \citep{Stickel_2007}.  Neither map has full coverage, with
MIPS missing data on small scales and ISOSS on larger.

We observed Cas~A using the Far-Infrared Surveyor
\citep[FIS;][]{Kawada_2007} instrument on-board AKARI. We obtained
fully-sampled images of sub-arcminute resolution covering a wide area
surrounding Cas~A in four photometric bands from 50 to
180\,$\mu$m. This is the wavelength range over which the emission from
newly formed hot dust becomes faint and emission from cold dust begins
to dominate. Therefore, these AKARI FIS images are very useful for
investigating the presence of colder components of dust in the
remnant.

The Balloon-borne Large-Aperture Submillimeter Telescope\footnote{\tt www.blastexperiment.info} \citep[BLAST;][]{Pascale_2008} was also used to observe the Cas~A region at 250, 350, and 500\,$\mu$m. These bands fill in the cold dust spectral energy distribution (SED) on the long wavelength side of the peak. The high mapping speed of BLAST means it was possible to cover a large area surrounding the SNR, giving data for both the SNR and the interstellar cloud structure in the surrounding region.

By contrast, the prior SCUBA maps are of a small area and
involve deconvolution of a three-beam chopping pattern which
reconstructs large scale power poorly.
AKARI and BLAST have the advantage that, like \textit{Spitzer}, they
are not required to perform chopped observations. As a result, our new
maps are sensitive to the large scale structure present in the Cas~A
field.
The relatively large area of these maps, coupled with their
sensitivity and wavelength coverage,
make them ideal for investigation of cold dust emission from Cas~A
and the interstellar clouds.

We describe the AKARI and BLAST observations in \S~\ref{data}. Several
distinct sources of emission are distinguishable in these maps.
Taking advantage of the AKARI spatial resolution, in
\S~\ref{internal_dust} we identify a new morphologically compact
source of ``tepid'' ($\sim$35\,K) dust emission centered on the SNR whose spectrum is also distinct, peaking between the SED peak of the hot dust in the SNR and that of the cold dust.
In \S~\ref{photometry} we perform aperture photometry on the maps in
the region of the Cas~A SNR.  The resulting global SED further illustrates the
different spectral components and shows that without the additional morphological information it is not possible to unambiguously distinguish the tepid dust component.
In \S~\ref{comp_sep} we fit the six-band AKARI-BLAST data with a
simple spectral model to make column density and temperature maps for
the cold dust.  These clearly illustrate the confused nature of cold
dust emission on the line of sight to the SNR.
In \S~\ref{foreground} the derived column density on the line of sight
is compared to that obtained by other techniques.
We present our conclusions in \S~\ref{conclusions}.


\section{Data}\label{data}

\subsection{AKARI}\label{akari_data}

The AKARI FIS observations of Cas~A were obtained on 2007 July 19 (Obs
ID: 1402802). FIS has four photometric bands at 65, 90, 140, and
160\,$\mu$m; two wide bands, 90 and 140\,$\mu$m, and two narrow bands,
65 and 160\,$\mu$m \citep{Kawada_2007}. FIS carried out a `slow-scan'
observation of Cas~A with a scan speed of 15\arcsec~s$^{-1}$ and a
reset interval of 1\,s. The size of the detector pixels was
$26\farcs8$ for the 65 and 90\,$\mu$m bands and $44\farcs2$ for the
140 and 160\,$\mu$m bands, approximately equal to the beam size. The
area covered for Cas~A is 40\arcmin\ $\times$ 12\arcmin. During the
observation, the dark signal was measured five times, at the
beginning, at the turning points of each scan leg, and at the end of
the observation, by closing a cold shutter. The variation of
responsivity was monitored by using an internal calibrator.

The raw data were processed with the official FIS slow-scan toolkit
\citep{Verdugo_2007}. The removal of bad data as well as cosmic-ray
hits and the correction of integration ramps were made in the
pipeline. During the responsivity variation correction, dark current
subtraction and flat fielding were performed. Since the dark level for
160\,$\mu$m was overestimated due to a large change of responsivity,
the 160\,$\mu$m data were excluded in this paper. The sigma-clipping
method was applied to calibrated signals within the size of a detector
pixel in order to remove small undetected glitches as well as other
artifacts. The image was reconstructed with the assumption that a
pixel value represents the uniform intensity over the pixel surface in
the fine image grid. The average value was taken from the multiple
values in the image grid. The absolute calibration was accomplished by
comparison with data from the Diffuse Infrared Background Experiment
\citep[DIRBE;][]{Silverberg_1993}. For point-source calibration, the
uncertainties of flux calibration are no more than 20\% for the two
short wavelength bands, and 30\% for the 140\,$\mu$m band
\citep{Kawada_2007}.

\subsection{BLAST}\label{blast_data}
BLAST made a `cap' observation \citep{Pascale_2008} of Cas~A during its first long duration balloon flight from Kiruna, northern Sweden, in June 2005. Maps were made with a slew rate of 6' s$^{-1}$ in azimuth, and a slow drift in elevation which produced a scan line spacing of $3\farcm25$ (about half an array width). Due to flight time limitations, it was only possible to perform 1.25 complete scan maps. The final images are $\sim$40\arcmin~$\times$~60\arcmin\ in size. The maps were made with the SANEPIC algorithm \citep{Patanchon_2008} and were calibrated using the procedure discussed in \cite{Truch_2008}.

Although the mirror diameter was 2\,m, the BLAST05 SANEPIC maps had only 3\arcmin\ full-width at half-power resolution due to an anomalous telescope beam, corrupted by some uncharacterized combination of mirror distortion and de-focus \citep{Truch_2008} (this was corrected for the BLAST06 flight).  Nevertheless, these maps have high signal-to-noise and are oversampled with 20\arcsec\ pixels, so that Lucy-Richardson (L-R) deconvolution can be used to improve the resolution significantly (Roy et al. 2009, in preparation).  After 32 iterations, the L-R maps used here have point spread functions of $1\farcm3$, $1\farcm6$, and $1\farcm9$ in the 250, 350, and 500\,$\mu$m bands, respectively, determined by measuring the size of a point source in the maps.  While greatly improved, these are still not as small as the nominal BLAST values of $0\farcm5$, $0\farcm66$, and 1\arcmin, or the even smaller values anticipated with the \textit{Herschel Space Observatory}\footnote{\tt http://www.esa.int/SPECIALS/Herschel/index.html}.

The AKARI maps have a well-established zero point, but the BLAST
SANEPIC processing filters the lowest spatial frequencies and produces
maps with zero mean value. In order to model the column densities and
temperatures in \S~\ref{comp_sep}, we need to establish the true zero
points of each BLAST map.  As described below, we use DIRBE maps of
the region to establish the appropriate mean value of the 250\,$\mu$m
BLAST map, and we use correlations between the BLAST maps to establish
the zero points at the two longer BLAST wavelengths. In practice these
two steps are done simultaneously.

\begin{figure} 
\includegraphics[scale=0.45]{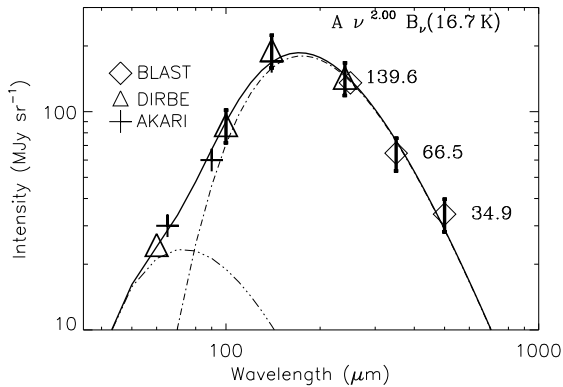}
\caption{Mean values of the maps within a DIRBE beam. 
The DIRBE photometry used to set the zero-points in the BLAST maps is shown as triangular data points, and the relative amplitudes of the BLAST mean values, as described in the text, are shown as diamonds. The solid curve is the sum of a cold dust curve with emissivity index $\beta=2$ and adjustable temperature. Also shown is a curve for non-equilibrium dust emission from small grains with $\beta=2$, $T=40$\,K and an amplitude 14\% of that of the cold dust.  The amplitude and temperature of the cold dust is chosen to minimize the deviation of this sum from the data points at the locations of the vertical error bars. The size of those error bars shows the assumed constant relative uncertainty corresponding to $\chi^2_\nu=1$ for the best fit. The pluses show mean values from the AKARI maps well away from the Cas~A SNR, and the agreement is excellent. The inferred mean zero-point values appropriate to the three BLAST bands are written next to each point.}
\label{fig:zpoints}
\end{figure}

Although the DIRBE beams are $1.3\times 10^{-4}$\,sr, much too large
to resolve the structure around Cas~A (there is of order one beam in the entire BLAST map), the experiment was explicitly
designed to measure total flux and is therefore ideal for establishing
the absolute intensity in each map. For the Cas~A position the DIRBE
intensities\footnote{\tt
  http://lambda.gsfc.nasa.gov/product/cobe/browser.cfm} are 24.5,
86.8, 191.2 and 142.9 MJy\,sr$^{-1}$ at wavelengths of 60, 100, 140
and 240\,$\mu$m. These values are plotted in Figure~\ref{fig:zpoints}.

The morphology of the structure in the BLAST maps, shown in Figure~\ref{maps}, are generally similar to
each other, indicating that the SED of the interstellar dust is
comparatively uniform across the region. The Pearson correlation coefficients
of the 350 and 500\,$\mu$m maps to the 250\,$\mu$m map are $r=0.98$
and $0.97$, respectively. The measured slopes, $A$, for the relation
$\hat I_\lambda = A\hat I_{250}$ are 0.51 and 0.28, respectively, for
$\lambda= 350$ and $500~\mu$m, where BLAST map intensities, $\hat
I_\lambda$ have zero mean and have been measured on maps with the
region of the Cas~A SNR masked out.
In the simple picture that the intensity variation within the maps is
dominated by variations in column density from place to place, these
linear relations should extrapolate to the origin if the maps had
correct zero points, and so knowledge of the absolute intensity in any
one BLAST band would allow inference of the zero points of the other
two maps.

As described in \S~\ref{comp_sep}, however, there is a mild
anti-correlation of temperature and column density: the dust in the
dense arms of the molecular cloud is approximately 2\,K cooler than
the dust in the less dense gas surrounding them. This biases the
measured slopes upward by 7\% and 11\% at 350 and 500\,$\mu$m.

The BLAST slopes, after correction for this small bias, are also
plotted in Figure~\ref{fig:zpoints}, relative to the 250\,$\mu$m point
on the solid curve.  This curve was obtained by fitting to the DIRBE
fluxes at 100, 140, and 240\,$\mu$m and the relative BLAST amplitudes
at 350 and 500\,$\mu$m, minimizing $\chi^2$ while varying the
amplitude of the curve and the temperature of the cold dust. We assume
the same fractional uncertainty in each band. The solid error bars in
the figure denote the bands used in the fit, and the size of the error
bars has been adjusted post-fit to show the fractional uncertainty
which would produce a reduced $\chi^2_\nu=1$.
The cold dust temperature was found to be 16.7~K. 
In order to accommodate the 60\,$\mu$m data, the curve includes a
contribution of non-equilibrium emission
\citep{Desert_1990,Li_and_Draine_2001} from very small dust grains
(VSGs) at an assumed temperature of 40~K whose peak amplitude has been
set to be 14\% of that of the cold dust component. However, the 60
$\mu$m data are not used in the fitting procedure. Both the cold dust
and the non-equilibrium components are assumed to have power law
emissivities, $\epsilon\propto \nu^\beta$ with $\beta=2$. Changing
$\beta$ to 1.5 does not improve the quality of fit, nor does it alter
the BLAST zero points by more than a few percent.

As a check, after the above fitting procedure was complete, the
results were compared to mean values from the AKARI maps, which are
shown as pluses in Figure~\ref{fig:zpoints}. The agreement is
excellent. The BLAST maps presented in Figure~\ref{maps} have had the
correct zero points applied.

\subsection{Maps}\label{textmaps}

\begin{figure*}
\includegraphics[scale=0.6]{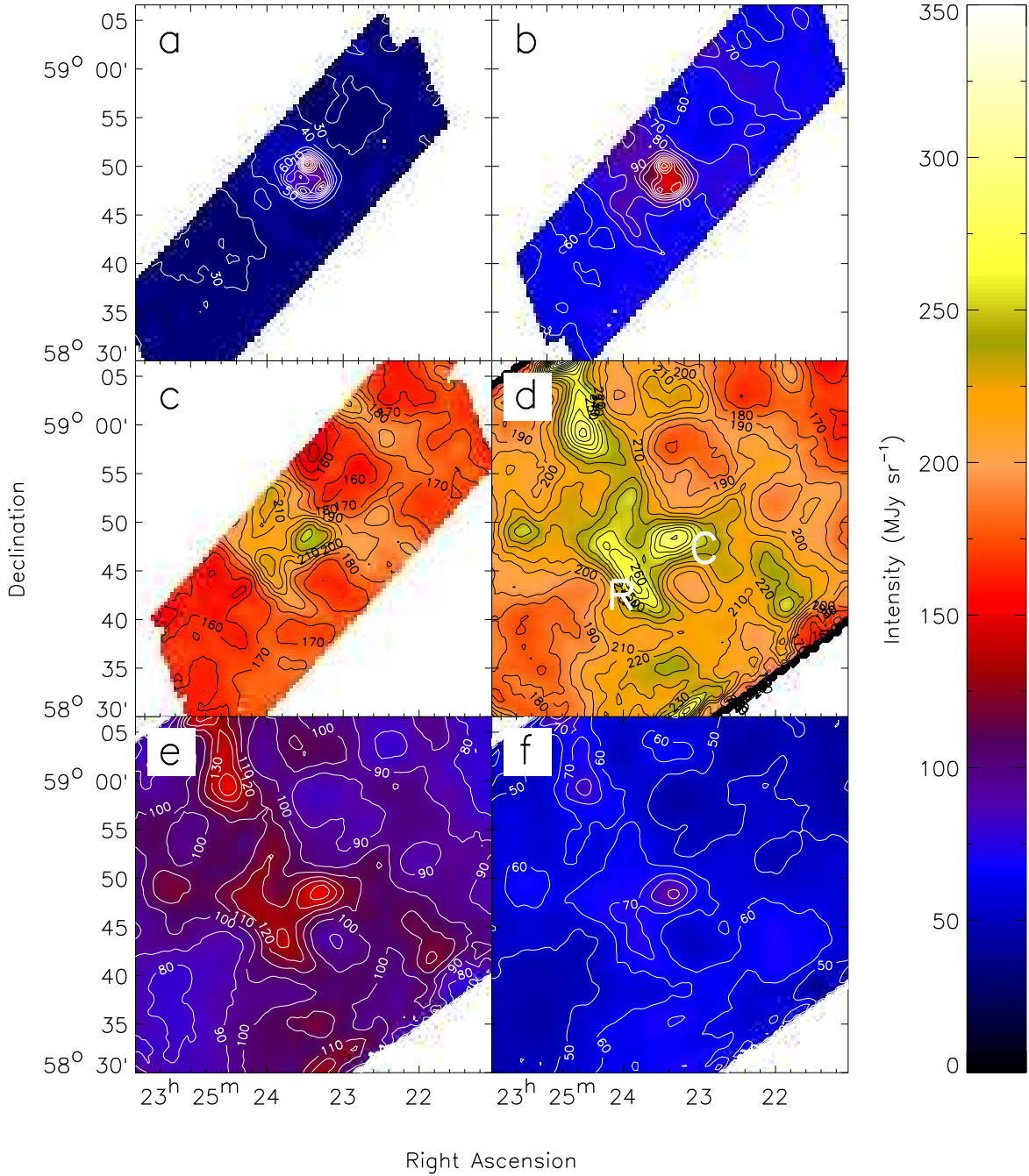}
\caption{AKARI 65, 90, and 140\,$\mu$m (a--c) and BLAST 250, 350, and 500\,$\mu$m (d--f) maps of the Cas~A SNR and surrounding ISM, brought to a common resolution of $1\farcm9$ and centered on the SNR. All maps are displayed with the same intensity scale. Contours are at the native resolution, in steps of 10\,MJy\,sr$^{-1}$.  The letters C and R label the line of sight elliptical ``cloud'' and ``ridge'' features described in the text above.}
\label{maps}
\end{figure*}

Figure~\ref{maps} presents the AKARI 65, 90, and 140\,$\mu$m and BLAST
250, 350, and 500\,$\mu$m maps.  All maps are centered on the same
location, 
are smoothed to a common resolution of $1\farcm9$, 
and are displayed with the identical range of brightness.

\begin{figure} 
\includegraphics[scale=0.45]{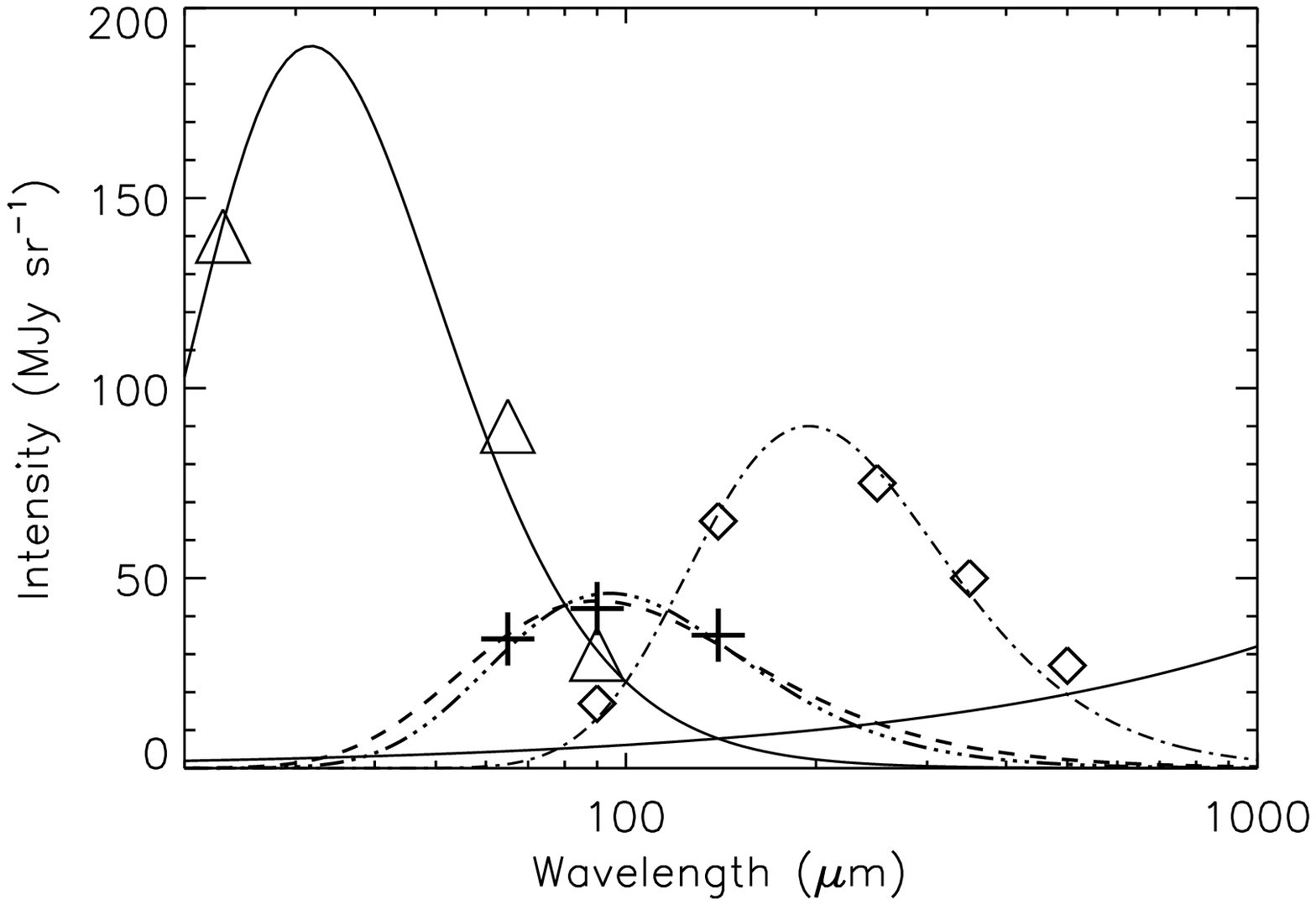}
\caption{Illustrative SEDs of four distinguishable sources
  of emission (not co-spatial).
The ``hot'' dust spectrum, shown as triangles, is the intensity in each map taken at the location of the peak of the 65 $\mu$m map, $\alpha=23^h23^m.4$, $\delta=58^\circ50^\prime.1$ (J2000) minus the local background measured at $23^h23^m$, $+58^\circ45^\prime$.  The illustrative curve near these points is a $\beta=2$ modified blackbody at $T=92$\,K.
``Cold'' dust (diamonds) is represented by the intensity of the ``ridge'' in the interstellar cloud structure to the south-east at $23^h50^m$, $+58^\circ44^\prime$, minus the respective map average. A $T=15$\,K, $\beta=2$ modified black body is drawn to guide the eye.
The pluses show the intensity of a ``tepid'' dust Gaussian central contribution, determined in \S~\ref{internal_dust}.  The two curves shown ($\beta=1$, $T=41$~K, dashed; and $\beta=2$, $T=33$~K, dot-dashed) fit the data equally well.
The lower solid curve shows the synchrotron spectrum measured at $23^h23\fm2$, $+58^\circ48\farcm8$, the brightest peak in the west of the SNR in the 86~GHz map by \cite{liszt_lucas_1999}, and extrapolated using spectral index $-0.72$.}
\label{fig:components}
\end{figure}

The relative intensity of emission from the SNR and interstellar
clouds varies significantly across the six bands. In the 65\,$\mu$m
image (Figure~\ref{maps}a) the SNR clearly dominates the map, with the
cloud structure barely visible.  Hot dust emission arises in the shell
of the SNR where there is a reverse shock. This is evident in the
shortest wavelength maps in the bright sources along a ring at the
apparent SNR outer edge.  To illustrate the spectral signature, the
brightness at the location corresponding to the brightest
peak in the 65 $\mu$m map minus the local background outside the SNR
is plotted in Figure~\ref{fig:components}. 
A point at 24\,$\mu$m is included to suggest the many short-wavelength
data that are available (\S~\ref{photometry}). When modeled as a
simple single-temperature modified blackbody, the hot dust temperature
$T_{\rm{H}}$ is $\sim 90$~K ($\beta=2$).  Note that this is a great
simplification.  Adopting a lower $\beta$ would broaden the spectrum,
possibly accounting for an underlying intrinsic temperature
distribution.  Nevertheless, $T_{\rm{H}}$ is in the range found by the
more detailed models by \cite{Rho_2008}.

The SNR is also dominant in the AKARI 90\,$\mu$m image
(Figure~\ref{maps}b). However, the emission ceases to be confined to a
simple ring.  The brightness at the center of the SNR rises slightly
going from 65 $\mu$m to 90\,$\mu$m, which is not consistent spatially or spectrally with
emission from either the hot or the cold dust component described
below.  In \S~\ref{internal_dust} we present an analysis of the
higher resolution AKARI data to estimate the properties of this new
source of intermediate temperature ``tepid'' dust emission (see resulting SED in
Figure~\ref{fig:components}).

The brightness of the interstellar emission increases at the longer
wavelengths, demonstrating that the dust there is at a lower
temperature than in the SNR. This ``cold'' dust emission peaks between
the 140 and 250\,$\mu$m data (Figs.~\ref{maps}c and \ref{maps}d), and
following a cold greybody SED recedes in the two longest wavelength
maps (nevertheless, it is still detected with high signal to noise).
The morphology of the cloud structure, both peaks and valleys, is
clearest in the 140 -- 350\,$\mu$m maps (Figs.~\ref{maps}c --
\ref{maps}e). Near the peak of the cold dust SED, the SNR does not
stand out as a special feature in the map. 
There is a brighter cold cloud to the north-east in the extended BLAST maps.  Closer to the SNR, there is a bright ``ridge'' of interstellar emission to the south-east and another elliptical ``cloud'' beginning to the west outside the remnant and appearing to project across its face. These two features are labelled ``R'' and ``C'' in Figure~\ref{maps}d respectively. It is clear that any emission associated with a possible component of fresh cold dust within the SNR will be highly confused by line of sight interstellar emission; this particular interstellar cloud will be called the ``line of sight cloud''.
The temperature of the cold dust throughout the map is $\sim16$~K (\S~\ref{comp_sep}).  A representative SED is shown in Figure~\ref{fig:components} for the molecular cloud ridge to the south-east of the SNR.

The diffuse interstellar emission across the 65 -- 90\,$\mu$m maps is generally too strong to be modeled along with the long wavelength emission as a single-temperature modified blackbody.  It is attributed to the above-mentioned non-equilibrium emission by VSGs (\S~\ref{blast_data}).

Synchrotron emission from relativistic electrons accelerated in the
SNR has been mapped extensively, including in the millimeter
\citep{liszt_lucas_1999,Synch_Wright_1999}.  The synchrotron power-law
SED rises smoothly toward longer wavelengths, and so eventually
dominates over the falling dust spectrum.  This is already becoming
evident in our 500\,$\mu$m BLAST map, where the SNR has recovered in
relative brightness compared to the interstellar emission.  By
850\,$\mu$m (SCUBA), the synchrotron emission is dominant.
A representative SED is shown in Figure~\ref{fig:components} for the
synchrotron peak in the west of the SNR.  This is drawn for a
power-law spectral index $\alpha = -0.72$, near that found by
\cite{Rho_2003}. Spectral index variations across the face of the SNR
\citep{Rho_2003,Dunne_2009} and with frequency \citep{Atoyan_2000}
are possible, but not critical for the analysis here.


\section{Tepid Dust Interior to the SNR}\label{internal_dust}

As discussed above, there is well-documented hot ($\sim 100$\,K) dust
in a broken ring closely associated with the reverse shock and the
X-ray emitting plasma \citep{Rho_2008}. However, toward the core in
the ejecta that has yet to be overrun by the reverse shock there could
be cooler dust that can be detected only at longer wavelengths
\citep{Dwek_Werner_1981,Mezger_1986}. Its minimum temperature would be set by the
ambient radiation field and so it would be at least as warm as the
surrounding interstellar dust.

To track ``tepid'' or intermediate temperature dust which is neither
the hot SNR ring component nor cold interstellar emission, the optimal
spectral region is 90 -- 160\,$\mu$m. Indeed, using multi-wavelength
(60, 100, 170, 200 $\mu$m) ISOPHOT images, \citet{Tuffs_2005}
presented evidence for emission from such tepid dust in the un-shocked
ejecta in two ways: (i) the morphology of the 100\,$\mu$m map shows
more central emission than expected from the hot dust component which
dominates at 60\,$\mu$m, indicating a cooler centrally-peaked
component, and (ii) the 170/200 brightness ratio is peaked on the SNR
rather than extending with the band of interstellar emission well
beyond the SNR to the west, indicating an emission component
associated with the SNR that is warmer than the cold interstellar
dust. We have confirmed these qualitative findings using the ISOPHOT
pipeline-processed data from the ISO archive.

The AKARI angular resolution and wavelength coverage are ideally
suited for pursuing approach (i) to the next level, using a
``spectrum-informed clean'' technique. Like the 60\,$\mu$m ISO image,
the 65\,$\mu$m AKARI image is dominated by the ring component. This
can be isolated by subtracting the local background of about
55\,MJy\,sr$^{-1}$. This ``background-subtracted'' map, a hot dust
template, can then be scaled, following a hot dust SED, and
subtracted from the background-subtracted 90\,$\mu$m image.  Choosing
the SED so as to remove (i.e., ``clean'') the morphological signature
of the hot dust ring, the striking result is an extended residual
feature at 90\,$\mu$m centered at $23^h23^m29^s$,
$+58^\circ48^\prime33^{\prime\prime}$ (J2000).  To the accuracy
available here, this direction is probably indistinguishable from the direction
to the center of the SNR. The shape of the feature is well
approximated by a Gaussian of FWHM 2\arcmin\ which is substantially
larger than the beam (37\arcsec) though smaller than the SNR. The
central surface brightness of this residual is about 20
MJy\,sr$^{-1}$, which can be compared to the total of 133
MJy\,sr$^{-1}$ in this same direction. Following the same procedure
with the 60 and 100\,$\mu$m pair of ISO images produces a residual of
very similar size and position, with central surface brightness 27
MJy\,sr$^{-1}$. This is an important check, because the images are not
completely free of striping. This quantifies the emission component
identified by \citet{Tuffs_2005} by its relative morphology.

An interesting question is whether this extended Gaussian component
can be seen in the short wavelength images that have been used to
generate the hot dust template. Depending on the three-dimensional
geometry, there will be some hot dust emission projected on the
central line of sight, though a peaked Gaussian arising in this way
would seem unusual. Furthermore, at 60 and 65\,\micron\ the surface
brightness toward the center is not only brighter than the background,
but compared to the hot dust ring is relatively brighter than is the
case at shorter wavelengths such as in the archival MIPS
24\,\micron\ image (convolved to the same resolution).
This is also the conclusion of \citet{Rho_2008} based on the MIPS
70\,\micron\ image \citep{Hines2004}.
Therefore, to probe this more probable alternative, we assume that the
central emission in the background-subtracted 65\,\micron\ map can be
attributed entirely to this Gaussian, which then would have a peak
brightness of 33\,MJy\,sr$^{-1}$.

\begin{figure}
\includegraphics[scale=0.5]{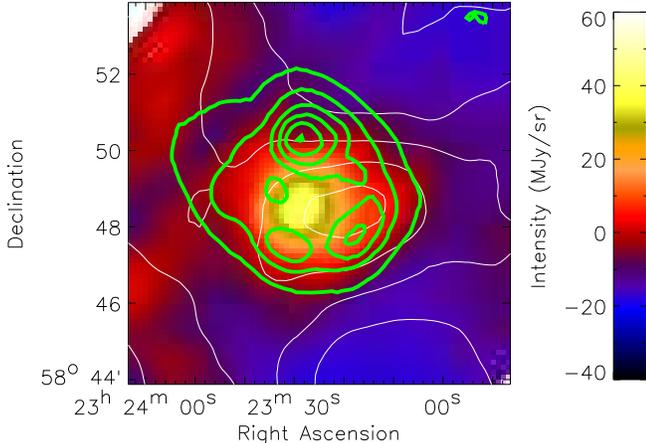}
\caption{AKARI 90\,\micron\ background-subtracted map of the Cas~A SNR
  from which the hot dust emission has been removed (``cleaned'').
  Heavy contours from the AKARI 65\,\micron\ map show the position of
  the hot dust ring and the thin contours from the BLAST
  250\,\micron\ map show the cold dust emission including the ``line
  of sight cloud'' and the ``ridge'' to the south-east.}
\label{residual}
\end{figure}

Subtracting this Gaussian component from the background-subtracted
65\,\micron\ map produces a new hot dust template with a larger dip in
the middle. Therefore, when this is subtracted from the
90\,\micron\ map, the Gaussian residual is still there, but with a
larger amplitude, 42\,MJy\,sr$^{-1}$ (see
Figure~\ref{residual}). Analysis of the 60 -- 100\,\micron\ ISO pair
yields a similar result. We consider this to be an upper limit.  The
60 -- 65 to 90 -- 100\,\micron\ color temperature is about 33\,K
($\beta = 2$) or 41\,K ($\beta = 1$).

Proceeding with this ``spectrum-informed cleaning'', now removing both
the hot dust and tepid Gaussian components from the 90 or 100\,$\mu$m
images, we find toward the SNR only a fainter residual in the west,
(coincidentally) at the projected distance of the reverse shock, and
extending to the southeast, a morphology crossing the remnant that is
qualitatively consistent with being interstellar. The latter
morphological component (called the ``line of sight cloud'' in
\S~\ref{textmaps} and below) can be seen directly in the 250 and
350\,\micron\ BLAST images (Figure~\ref{maps}), even without any
component subtraction, and in the cold dust column density map
(Figure~\ref{model_maps}b below in \S~\ref{comp_sep}).  As in the MIPS
160\,\micron\ image \citep{Krause_2004_CasA} and the long wavelength
ISO maps, this extends to the west well outside the SNR.

In the two intermediate-wavelength images (AKARI 140\,\micron\ in
Figure~\ref{maps}c; ISO 160\,\micron) the hot dust ring component is no
longer obvious. But in both there are two distinguishable peaks of
emission, toward the center and the band from the western peak
extending to the southeast. The Gaussian component could be as bright
as 35 MJy\,sr$^{-1}$ (with the second alternative hot dust template),
fairly consistent with extrapolation of the tepid dust SED from shorter
wavelengths. Compared to the situation at 100\,\micron, the residual
line of sight cloud projected on the SNR and the adjacent ``ridge'' of
interstellar emission to the south-east have become brighter, as
expected for cold dust.

The peak amplitude of this central Gaussian feature, from the three
maps where it can be distinguished, is plotted as pluses in
Figure~\ref{fig:components}. Two modified blackbody curves plotted
there ($\beta=2$, $T=33$~K and $\beta=1$, $T=41$~K) describe the
measured amplitudes equally well.  Both indicate only a small flux
from this component at longer wavelengths where the combination of
coarser BLAST angular resolution and comparatively intense line of
sight emission have not allowed direct measurement.

As mentioned above, hot dust emission is correlated with X-ray line
and continuum emission and infrared line emission from the shocked
gas.  It is therefore interesting to ask if there is any tracer of the
plasma that is coincident with the central tepid dust emission.
\cite{Smith_2009} show maps of several lines from different ions
tracing different physical conditions in the SNR.  The [Si II]
34.8~\micron\ emission in their Figures~2 and 7 is from the unshocked
ejecta in the interior of the SNR inside the reverse shock boundary
(see their Figure~4).  The peak position of this central emission
appears to be consistent with the location of our Gaussian component.
L.~Rudnick (private communication) finds from a map of the central low
density areas convolved to the AKARI resolution that the FWHM is
comparable too.

Identification of this tepid dust Gaussian component clearly depends on
clues from how the SNR morphology changes with wavelength.  Using the
higher-resolution multi-wavelength morphological information
anticipated from \textit{Herschel}, it will be very interesting to
characterize this component in more detail for comparison with such
(3-dimensional) tracers.

\subsection{Mass of Tepid Dust}\label{coolmass}

Dust emission maps record surface brightness $I_\nu$ which in turn
depends on the dust mass column density $M_d$ through
\begin{equation}
M_{\rm{d}} = I_\nu/\kappa_\nu B_\nu(T_{\rm{d}}),
\label{column}
\end{equation}
where $\kappa_\nu$ is the dust mass absorption coefficient, and
$T_{\rm{d}}$ is the temperature of the particular dust component.
Since dust in this Gaussian component is assumed to be associated with
the SNR, at distance $D$, integration over the face of the remnant
results in
\begin{equation}
\mathcal{M}_{\rm{d}} = S_\nu D^2/\kappa_\nu B_\nu(T_{\rm{d}}),
\label{mass}
\end{equation}
where $S_\nu$ is the measured flux density and $\mathcal{M}_{\rm{d}}$ is the
derived dust mass.

Adopting a central surface brightness 40 MJy\,sr$^{-1}$ at 100\,\micron, FWHM 2\arcmin, and $T_d = 33$~K, we get a dust mass $0.055 \times (30\, {\rm cm^2 g^{-1}})/\kappa_{\rm{100}} $\,M$_{\odot}$.  It would be a factor two lower for $T_d = 41$~K.  As is the case for the mass estimates for other dust components, the other large source of uncertainty comes from $\kappa_{\rm{100}}$.  The normalizing value used here at 100~\micron\ is appropriate to interstellar dust \citep{Li_and_Draine_2001} but the actual value might reasonably be different by a factor three. 

The derived mass for this new tepid component is comparable to that estimated previously for the hot dust component by \cite{Rho_2008} (see \S~\ref{The_CasA_SNR}).  Following their arguments and caveats, such dust yields could contribute to the dust masses seen in high redshift galaxies, but they are less than the required level 0.4-1\,M\subsun~estimated by \citet{Dwek_2007} to account for the dust seen at high redshift.  However, when combined with the previously reported hot and cold dust masses this makes supernova remnants a more plausible source of this dust.


\section{Global Spectral Energy Distribution}\label{photometry}

We measure the spatially-integrated (``global'') flux density $S_\nu$ in the direction of Cas~A using a
simple aperture photometry method. A three-arcminute radius aperture
is used to measure the SNR flux, and an annulus with inner and outer
radii of 4\arcmin\ and 5\arcmin\ respectively is used to measure the
background. Only information from the western side of the annulus is
used to estimate the background, as the eastern side contains elements
of large scale cloud structure, which would result in an overestimate
of the background flux.

\begin{figure}
\subfigure[]{\label{SED1}
\includegraphics[scale=0.45]{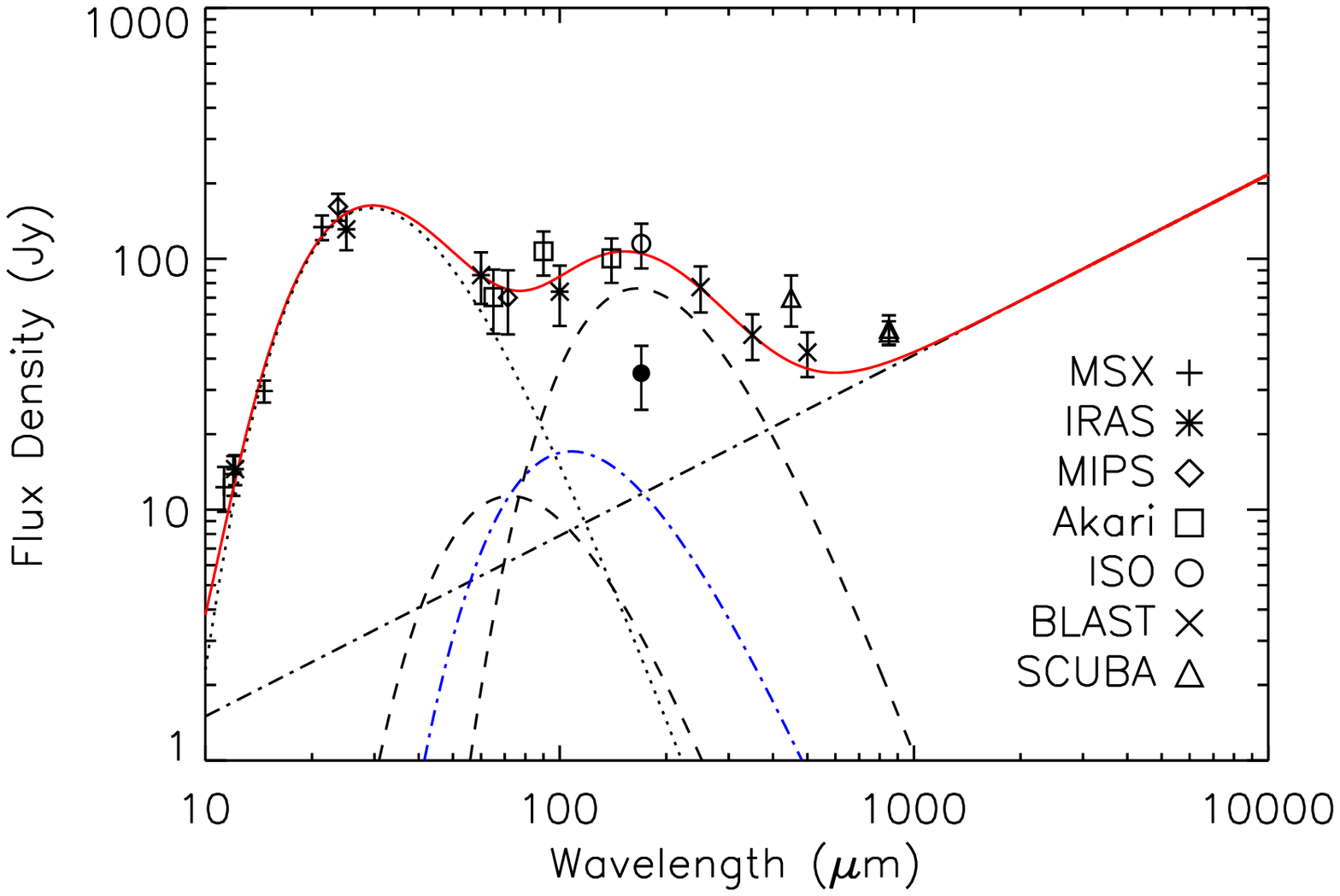}}
\subfigure[]{\label{SED2}
\includegraphics[scale=0.45]{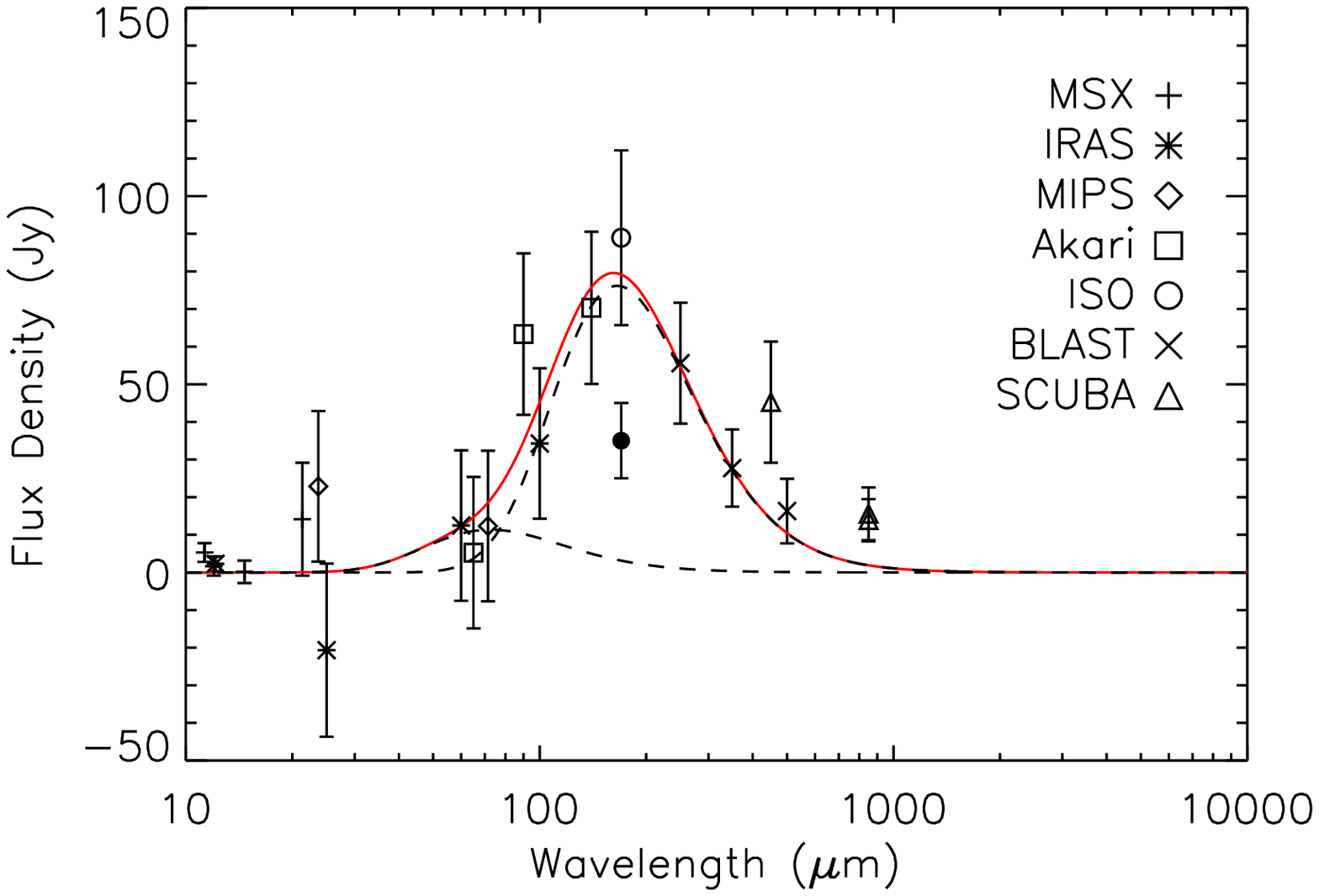}}
\caption{Global SED of the Cas~A SNR. 
A fit was made to published MSX, IRAS, and Spitzer/MIPS photometry \citep{Hines2004} together with the new AKARI and BLAST and remeasured ISO 170\,$\mu$m (open circle) data, but neither the original \cite{Tuffs_1999} value (filled circle) nor the SCUBA data (triangles) were used.  The model used for this fit is given in equation~\ref{eq_sed_components}, and contains physically motivated components attributable to line of sight interstellar dust, and dust and synchrotron emission from the SNR.  The two broad peaks in panel (a) are from hot SNR dust (dotted curve) and cold dust (upper dashed curve) with derived temperatures 99$\pm$3\,K and 16.6$\pm$1.3\,K, respectively.  More minor components are the new tepid dust Gaussian component in the SNR (\S~\ref{internal_dust}; dot-dash curve) and VSGs (lower dashed curve).  These contribute in the spectrally-confused 100\,\micron\ region and cannot be separated using this global SED alone.  
For panel (b), the flux of the synchrotron and hot and tepid dust
components have been subtracted to highlight the contribution of cold
dust and, less importantly, VSGs.}
\label{sed}
\end{figure}

In addition to measuring the flux density of Cas~A in the six bands
65, 90, 140, 250, 350, and 500\,$\mu$m from AKARI and BLAST, we
remeasured the Infrared Space Observatory \citep[ISO;][]{ISO_1996}
170\,$\mu$m photometry using data from the archive. We obtained a
value of 101$\pm$20\,Jy, significantly higher than the 35$\pm$10\,Jy
obtained by \cite{Tuffs_1999} which we believe was a lower limit. The
AKARI, BLAST, and ISO values are given in Table~1, and
plotted in Figure~\ref{sed} along with published MSX, IRAS and MIPS
data \citep{Hines2004}.

\begin{table}
\begin{center}
\begin{tabular}{cr@{~$\pm$~}l}
\hline
\hline
Wavelength ($\mu$m)    &       \multicolumn{2}{c}{Flux Density (Jy)}   \\
\hline
65      & ~~~~~71      & 20    \\
90      & ~~~~~105     & 21    \\
140     & ~~~~~92      & 18    \\
170     & ~~~~~101     & 20    \\
250     & ~~~~~76      & 16    \\
350     & ~~~~~49      & 10    \\
500     & ~~~~~42      & 8     \\
\hline
\end{tabular}
\caption{Table of Cas~A flux density measurements for the AKARI and
  BLAST bands, and the disputed ISO 170\,$\mu$m band. These were made using the
  aperture photometry described in (\S~\ref{photometry}).}
\end{center}
\label{fluxtable}
\end{table}

In order to fit these data we developed a physically motivated model
with four dust components as discussed above, including hot dust
(subscript H), the tepid Gaussian component (G), cold dust (C), and VSGs
(V), together with synchrotron emission, and described by:
\begin{equation}
S_\nu = \sum_{i=1,4} \Omega_i \nu^{\beta_i} B_\nu(T_i) + S_{\rm{0}}(\nu/\nu_{\rm{0}})^{\alpha},
\label{eq_sed_components}
\end{equation}
where the scale factors $\Omega$ are related to the
spatially-integrated column density or mass.  All four dust components
have been approximated as single-temperature modified blackbodies.  We
obtain the synchrotron fluxes by extrapolation from the map of
\cite{Synch_Wright_1999} and subtract them from $S_\nu$ and so these
data are not part of the fit to find the dust parameters.

For the cold dust (largely interstellar), we take $\beta_{\rm C}=2$ \citep{Li_and_Draine_2001} and find the other two parameters, $\Omega$ and $T$, from the fit. 

The value of $\beta_{\rm{H}}$ is much less certain.  Furthermore, this
is a highly simplistic model, as the hot modified-blackbody
underestimates the emission from polycyclic aromatic hydrocarbons
(PAH) and short-wavelength SNR features seen by \cite{Rho_2008}.
$\beta_{\rm{H}} = 1$ might be a better approximation to broaden the
model SED to accommodate the range of hot dust temperatures, but we
take it to be 2 to better describe the longer wavelength emission near
100\,$\mu$m (Figure~\ref{fig:components}). The hot dust emission from
the SNR dominates the total flux emitted in the short wavelength
region of the SED and so its characteristic temperature $T_{\rm{H}}$
is well determined, but discovering exactly how the SED extrapolates
to longer wavelengths requires morphological information, and is not
the focus here.

The single-temperature approximation is least appropriate for the non-equilibrium emission by VSGs which would have a range of temperatures, but this is not critical here since the VSG emission is relatively weak. To help constrain the model, we take a representative $T_{\rm{V}} = 40$~K, $\beta_{\rm{V}} = 2$, and $\Omega_{\rm{V}}/\Omega_{\rm{C}} = 1.8 \times 10^{-3}$ determined from the large scale interstellar regions of the maps (\S~\ref{comp_sep}). Consistent with the integrated spectrum shown in Figure~\ref{fig:zpoints}, this results in the peak of the VSG emission being about 14\% of that of the cold dust.

We take the parameters of the Gaussian component from our study in \S~\ref{internal_dust}, where the flux at 100~\micron, near the peak, is found to be 16~Jy and the SED is as shown in Figure~\ref{fig:components}.

Fitting the remaining parameters to the global spectrum, the total
model spectrum and its components are plotted in Figure~\ref{sed}.
The original ISO 170\,$\mu$m point and the SCUBA data were not
included in our fits.  The ISO flux we measure is consistent with the
new BLAST and AKARI data, as well as with previously published data at
other wavelengths. This fit does not appear consistent with the SCUBA
data at either 450 or 850\,$\mu$m, with the BLAST 500\,$\mu$m point
being $\sim$40\% (1.8$\sigma$) lower than the SCUBA 450\,$\mu$m
value. We cannot find any explanation for this disparity.

We find $T_{\rm{H}} = 99\pm$3\,K and $T_{\rm{C}}=16.6\pm$1.3\,K. For the same $\kappa$ as adopted for the tepid dust (for lack of better constraints), the independent masses of hot and cold dust (if the latter were all at 3.4~kpc) are 0.003 and 10\,M$_{\odot}$, respectively.  The latter is the potential amount of cold dust associated with the SNR, significant background emission having been subtracted when carrying out the photometry.  This limit, close to that estimated by \cite{Krause_2004_CasA}, illustrates just how difficult it is to separate out cold dust in the SNR, given a strong interstellar signal. This too needs to be addressed using morphological information.  It is noting that when the newly measured ISO point is exclulded from the fit there is only a negligible effect, changing the derived parameters by less than 1-$\sigma$ as compared to a fit performed without this datum.


\section{Component Separation}\label{comp_sep}

The photometry in Figure~\ref{sed} benefits from greater spectral coverage to provide an improved measurement of the spectral energy distribution toward Cas~A, but does not explicitly address whether that emission, particularly the cold dust, is line of sight interstellar emission or is intrinsic to the SNR. Here we exploit the six large area AKARI and BLAST maps over the region observed in common.  The maps were convolved to a resolution of $1\farcm9$ and re-gridded to the same pixelization to produce a coarse spectral ``cube''.  Using a multi-component model akin to equation~\ref{eq_sed_components} in \S~\ref{photometry}, an independent fit to the SED at each pixel in this cube is carried out and maps of each free parameter obtained.

We first subtracted the SNR synchrotron component from the data and
so, as in \S~\ref{photometry}, this model component is not a part of
the fit. We convolved the 83~GHz radio map of \cite{Synch_Wright_1999}
to produce a synchrotron template, and subtracted the appropriate
intensity by extrapolating the spectrum assuming a power-law spectral
index of $\alpha = -0.72$ (see \S~\ref{textmaps}).  This left us with
maps containing only thermal dust emission, but still four components
on the line of sight to the SNR, which introduces too many free
parameters.  Therefore, benefiting from the higher resolution analysis
in \S~\ref{internal_dust}, we also subtract from the maps the tepid
dust Gaussian component found there.

Because the analysis here uses only the AKARI and BLAST maps, it does not
contain information from wavelengths shorter than
65\,$\mu$m. Consequently it is not necessary or possible to have a
sophisticated modeling of short wavelength emission sources (e.g.,
PAHs and shorter wavelength VSG emission). We simply used $T_{\rm{V}}
= 40$~K and $\beta_{\rm{V}} = 2$ as above.  Additionally, toward the
SNR the intensity of the VSG component used in this simple model is
degenerate with the hot dust ring component and tepid dust Gaussian
component because of the overlapping SEDs, meaning that our analysis
is not sensitive to whether the ratio of VSG to cold dust emission for putative
SNR dust might be different than the equivalent ratio for the
interstellar line of sight cloud.  Away from the SNR we allowed this ratio
to vary in order to examine its statistical properties, and then we
adopted the average ratio for the confused line of sight to the SNR.
The properties of the cold dust (column density and temperature)
outside the SNR are quite similar whether this average ratio is used
in the model for the whole map, or is allowed to be fit.

The hot dust component should only be included in the model for pixels
where the SNR contributes to the intensity. In order to assess in an
objective automated way whether or not the data to which the model is fit
contain emission from the SNR, the AKARI 65 to 140\,$\mu$m ratio was
examined.  A maximum ratio of 0.22 was found for all regions beyond a
4\arcmin\ radius from the SNR.  Therefore, only when the ratio exceeded this
value was this hot component included in the model fit.  The
temperature of the hot component is held fixed at
100\,K, as derived in \S~\ref{photometry} with $\beta_{\rm{H}} = 2$,
and its amplitude (column density) allowed to vary.  As an
alternative, we experimented with removing this emission component
from the map using a scaled hot dust template. This final cleaning
step was not entirely satisfactory, probably because of changes in the
hot dust SED (temperature) across the face of the SNR, but again the
derived properties of the cold dust were not very sensitive to this
choice.  As another sensitivity test, we fit using a hot dust model
with $T_{\rm{H}} = 120$~K and $\beta_{\rm{H}} = 1$ and again the cold
dust parameters were robust, because the overlap of the cold and hot
SEDs is not great (Figure~\ref{sed}).

Our six-wavelength data are fit using a model which contains
either two or three free parameters, depending on whether the data
being fit contains a contribution from the SNR. These parameters are
the hot dust amplitude, 
the cold dust amplitude, and the cold dust temperature, $T_{\rm{C}}$.
Implicit in this is the assumption that the cold dust model component
is sufficient to describe the emission from both the SNR and line of
sight interstellar clouds.  A reduced $\chi^2$ measurement was
calculated for each pixel fit; the median value over the map was 0.89.

\begin{figure*}
\centering
\subfigure[Cold temperature]    {\label{model_temp}			 
\includegraphics[width=0.3\linewidth]{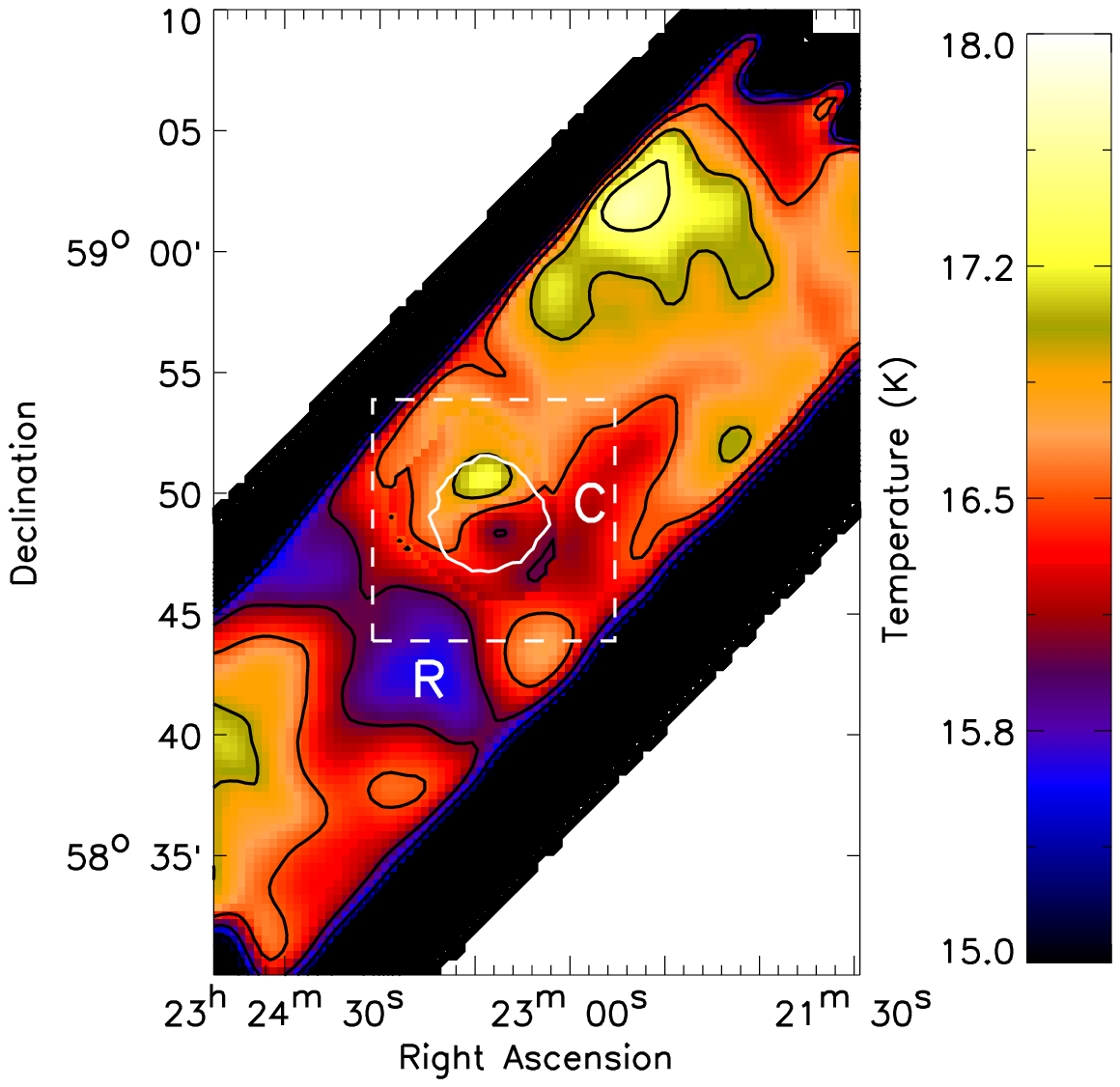}}
\subfigure[Cold column density] {\label{model_den1}			  
\includegraphics[width=0.3\linewidth]{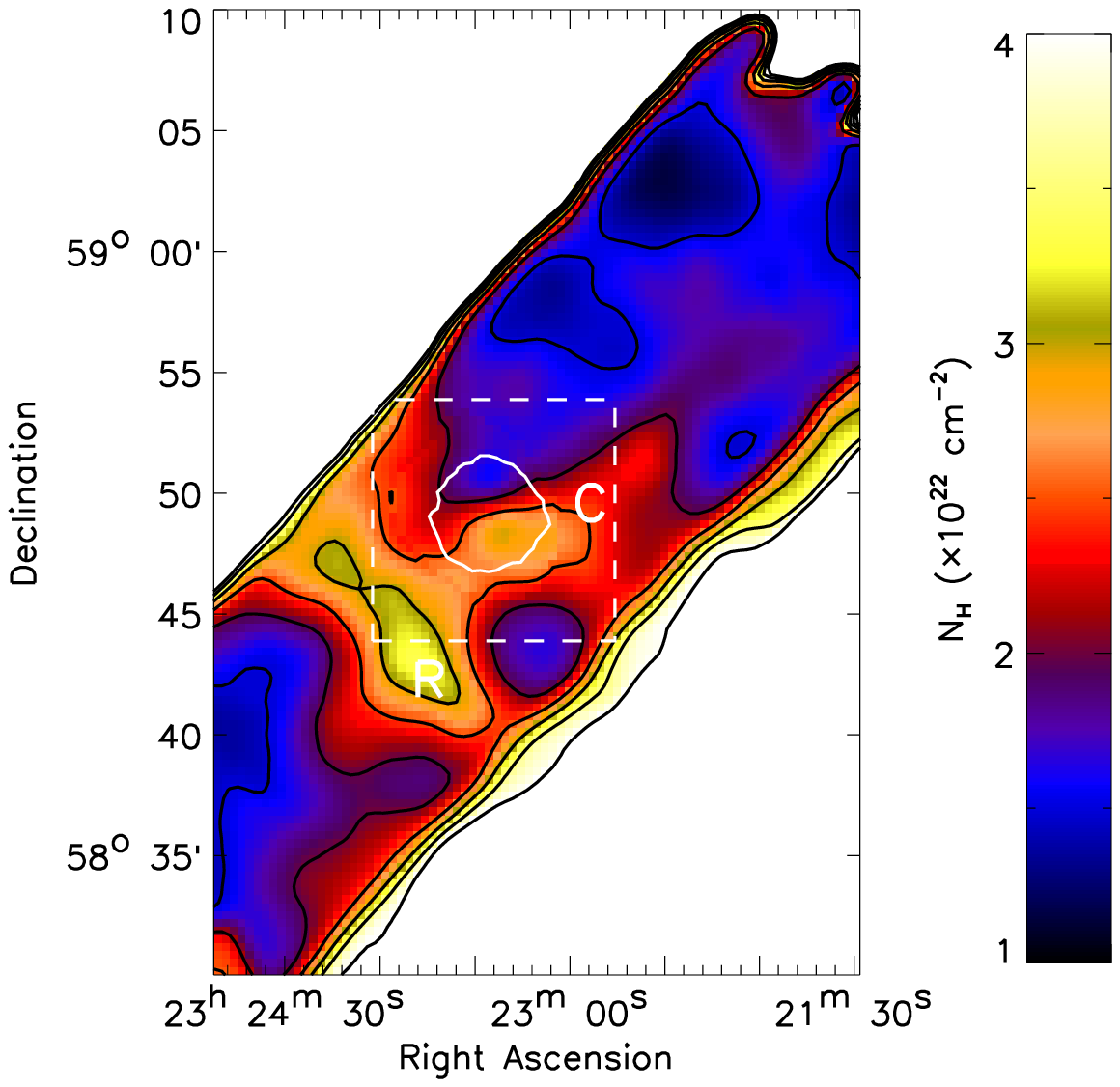}}
\subfigure[Hot column density]  {\label{model_den2}			 
\includegraphics[width=0.3\linewidth]{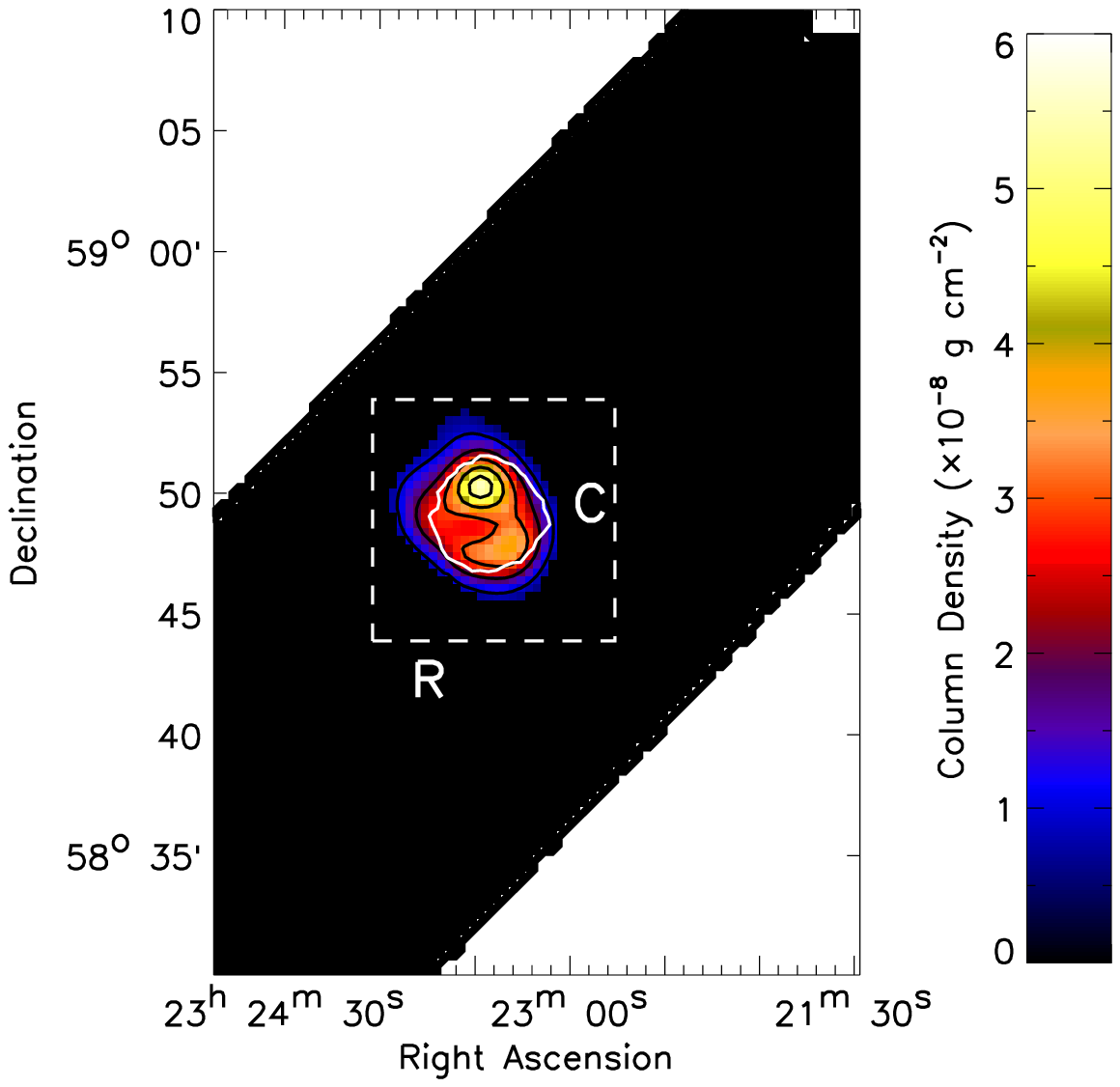}}
\subfigure[]{\label{model_temp_zoom}
\includegraphics[width=0.3\linewidth]{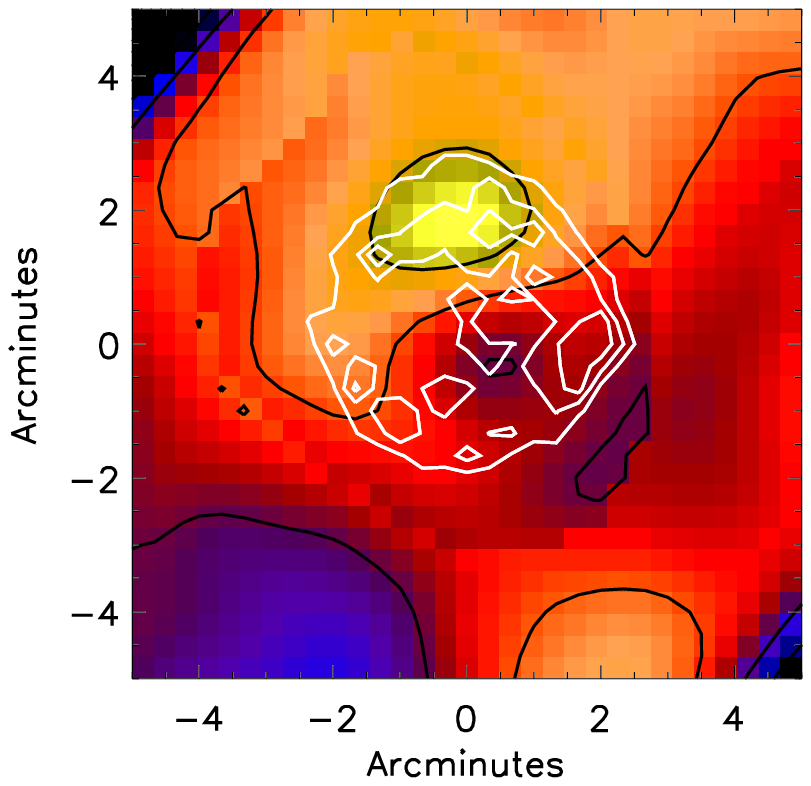}}
\subfigure[]{\label{model_C_dens_zoom}
\includegraphics[width=0.3\linewidth]{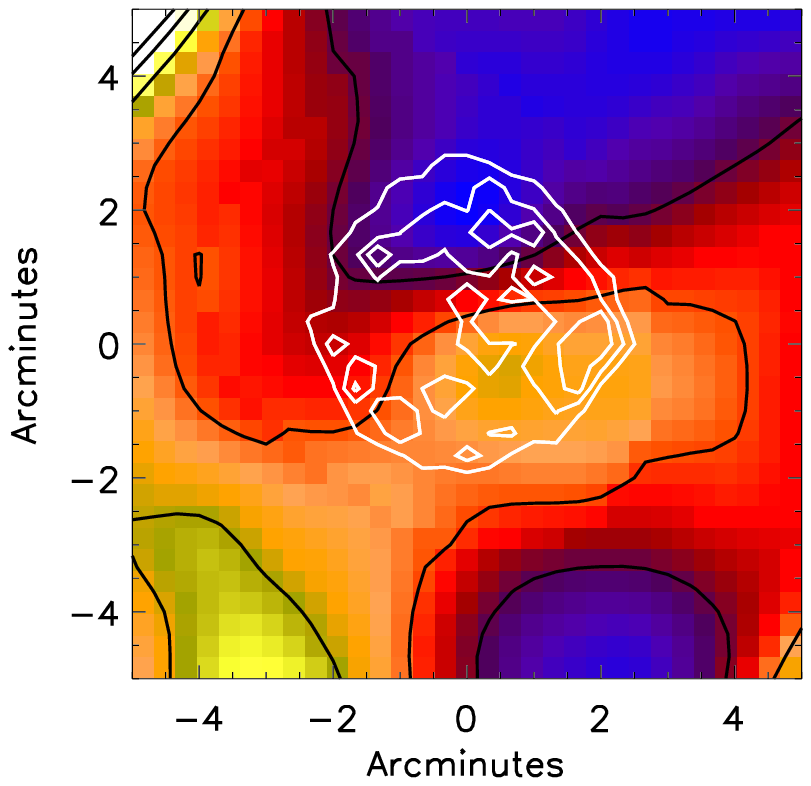}}
\subfigure[]{\label{model_h_dens_zoom}
\includegraphics[width=0.3\linewidth]{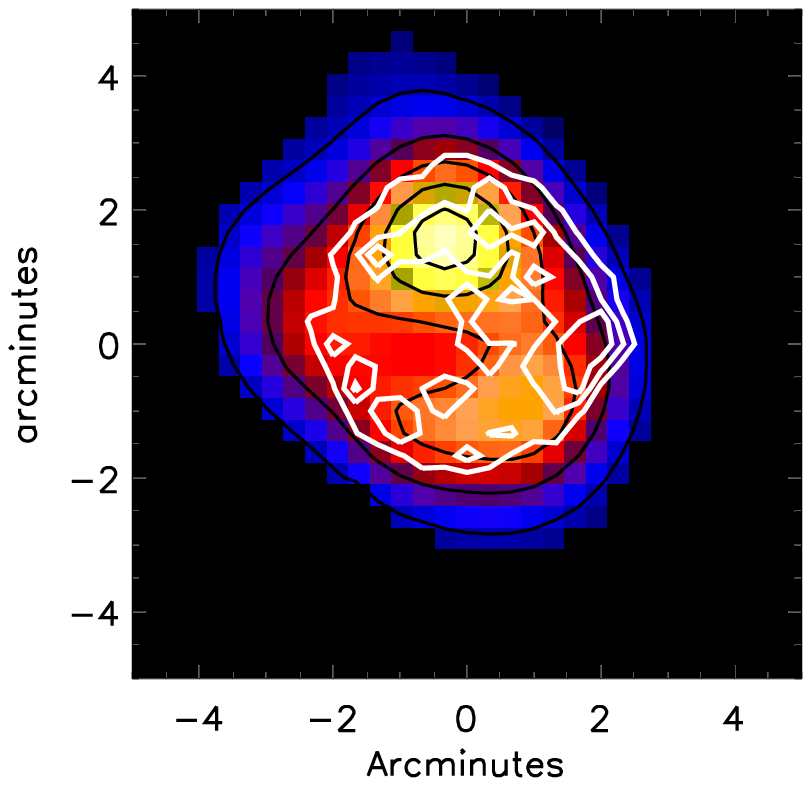}}

\caption{Output maps for the three free parameters of the spectral
  model: 
temperature of the cold dust and column density of both cold and hot dust. In the upper panels a single white contour from the 83\,GHz radio map \citep{Synch_Wright_1999} shows the approximate position of the Cas~A SNR forward shock \citep{Helder_2008}. The letters C and R label the elliptical line of sight ``cloud'' and ``ridge'' featured high-lighted in \ref{maps}d respectively.  The contour lines vary in steps of 0.5\,K, $0.5 \times 10^{22}$\,cm$^{-2}$, and 1$\times$10$^{-8}$\,g\,cm$^{-2}$ for maps a--c, respectively. Note that the outer edge regions are corrupted by convolution artifacts.
The lower panels show magnified versions of the output within the
white dashed square centered on $23^h23^m24^s$,
$+58^\circ48^\prime54^{\prime\prime}$, overlaid with several white
radio contours.}

\label{model_maps}
\end{figure*}

The output maps are presented in Figure~\ref{model_maps}, with
contours from the radio synchrotron emission \citep{Synch_Wright_1999}
overlaid to show the location and approximate size of the SNR.

We find interstellar cloud temperatures ranging between 15 and 18\,K,
with a median of
16.5\,K (see Figure~\ref{model_maps}a), in keeping with the temperature found by \cite{Wilson_1993} and the cold dust temperature found by \cite{Dunne_2003_CasA}.  The fine structure 
in this temperature map 
reflects some interplay between the various model components contributing to the emission model, but the main features there are robust against exactly how these spectral components are treated.

The Cas~A SNR does not stand out particularly.  There is a slightly hotter spot to the north of the SNR whose amplitude changes with the choice of alternative hot dust models and so probably reflects a deficiency there where the hot dust component is very bright. There is no telltale change in temperature at the location of Cas~A, to indicate the presence of a significant secondary cold dust grain population at a different temperature to the interstellar cloud.

The cold component column density map (Figure~\ref{model_maps}b) shows distinctly the cold dust ridge (R) to the south-east of the SNR, and the cold elliptical region (C) to the west of and overlapping the SNR (the ``line of sight cloud'').

\begin{figure}
\includegraphics[scale=0.5]{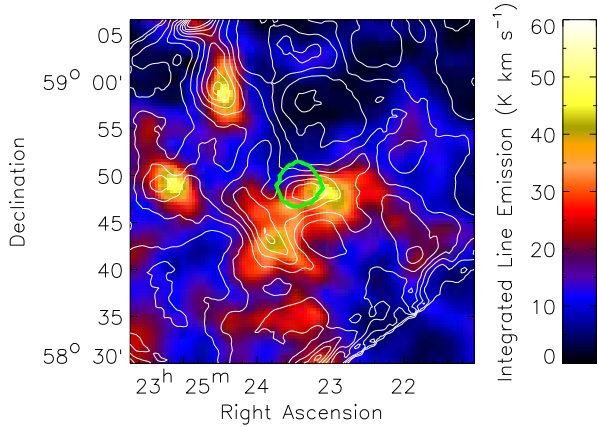}
\caption{Integrated $^{12}$CO~($1-0$) line emission from the FCRAO
spectral cube in the CGPS \citep{CGPS}. Thin contours from the BLAST 250\,\micron\ map show the cold dust emission including the ``line of sight cloud'' and the ``ridge'' to the south-east. Thick contour indicates the position of the radio SNR as in Figure~\ref{model_maps}.}
\label{12co}
\end{figure}

Note that feature C is not as strong as R, is not obviously correlated with any morphological features of the SNR, and extends to the west beyond the remnant.  
The cold column density map resembles what is seen in integrated
molecular line emission (Figure~\ref{12co}; see also Figure~4 of
\citealt{liszt_lucas_1999}, Figure~1 of \citealt{Krause_2004_CasA}, Figure~1 of \citealt{Wilson_Batrla_2005},
and
Figure~8 of \citealt{Dunne_2009}).
Therefore we conclude that there is, by chance, a substantial amount of interstellar material projected on the Cas~A line of sight.  This makes it difficult to make a statistically significant detection of cold dust distinctly associated with the SNR with these data.  It is likely that this confusion affected the measurements made by \cite{Dunne_2003_CasA} based on chopped SCUBA observations.  We can not identify the polarized dust source found by \cite{Dunne_2009}; however, based on our analysis of the data, this is to be expected as the flux density level predicted would be too faint to detect in our observations.

There is a large warmer-than-average region of $\sim$17\,K to the
north west which intersects with the remnant. Comparison with the
column density map shows this to be part of a general anti-correlation
between temperature and column density across the map.  This could
reflect a deficiency in the modeling, since the two are certainly
inversely coupled, non-linearly, through equation~\ref{column}.  But
it seems plausible that this has a physical origin.  The high column
density regions are molecular and, being more shielded from the
interstellar radiation field which heats the dust, would be cooler.
The low column density regions also tend to have a higher ratio of VSG
emission, when this is allowed to vary in the fit, though this might
be indicative of another effect as well, evolution of the grain size
distribution.

The third output map, Figure~\ref{model_maps}c, shows the column
density from the hot SNR dust component. The automated fitting has
invoked this component only at the location of the SNR,
the interstellar dust model being sufficient at locations away from
the SNR.  As expected, the structure of the emission is similar to
that in the 65\,$\mu$m band when convolved to the same resolution as
the BLAST data.

Quantitatively, the hot and cold component amplitudes were converted
to dust mass column density using equation~\ref{column}.  In principle
different values of $\kappa_\nu$ would be used for the different dust
components in different environments but, in the absence of persuasive
evidence, we proceeded as in \S\S~\ref{internal_dust}-\ref{photometry}
and adopted those of \cite{Li_and_Draine_2001} ($\kappa_{\rm{100}} =
30\, {\rm cm^2 g^{-1}}$; $\beta = 2$) which are suitable for the
diffuse ISM.  Since much of the ISM in this region is in molecular
form, these might not be entirely appropriate even for that component.
The value of $\kappa_{\nu}$ represents the greatest uncertainty in
this analysis, with published values spanning two orders of magnitude,
depending on the dust type and environment \citep{Dunne_2003_CasA}.

For the cold dust component, which is dominated by the line of sight interstellar dust, we further convert the dust mass column density into $N_{\rm{H}}$ by dividing by $1.9 \times 10^{-26}\, {\rm g\, H^{-1}}$ \citep{Li_and_Draine_2001}. $N_{\rm{H}}$, used in Figure~\ref{model_maps}b, is the metric quantifying X-ray photoelectric absorption, discussed in \S~\ref{foreground}.  The column densities are large throughout the map, of order $10^{22}$~cm$^{-2}$.  Similarly, this can in turn be converted to optical extinction $A_{\rm{V}}$ by dividing by $1.9 \times 10^{21}\, {\rm cm^{-2}\, mag^{-1}}$ \citep{Bohlin_1978}, equivalent to multiplying the mass column density directly by $2.8 \times 10^4\, {\rm cm^2\, g^{-1}\, mag}$.  The relationship between $N_{\rm{H}}$ and $A_{\rm{V}}$ is empirically calibrated only below about $N_{\rm{H}} = 0.4 \times 10^{22}\, {\rm cm^{-2}}$ or $A_{\rm{V}}= 2$ \citep{Kim_1996} but might still be a reasonable approximation for the interstellar material spread out along this long line of sight.

Note that both $N_{\rm{H}}$ and $A_{\rm{V}}$, being scaled from estimates based on optically-thin submillimeter emission, measure the total column extending through the Galaxy.  The column in the foreground of Cas~A should be comparable to this, simply given the large distance to the SNR and Galactic latitude, $-2.1$\deg, and longitude, 111.7\deg. In fact, the velocity range of molecular lines seen in absorption against Cas~A is the same as that seen in emission (\S~\ref{foreground}) suggesting that Cas~A is beyond most of the Perseus arm gas. Judging from the \ion{H}{1} 21-cm emission-line spectrum along adjacent lines of sight in the CGPS data cube \citep{CGPS}, up to $0.1 \times 10^{22}$~cm$^{-2}$ might be in the background, or less than 10\%.

If the cold dust projected on the remnant were all at that distance, the mass column density integrated over its face in a 5\arcmin\ diameter aperture would amount to $\sim 40$~$M_\odot$.  This is a few times larger than found in \S~\ref{photometry} because here there has been no subtraction of a ``background''; Figure~\ref{model_maps}b shows that the column density on surrounding interstellar lines of sight is substantial.  \cite{Dunne_2009} suggest detection of 1~$M_\odot$ in the SNR whereas \cite{Krause_2004_CasA} find an upper limit of 0.2~$M_\odot$.  One has to beware of the different $\kappa$ and $T_{\rm{C}}$ that have been adopted in different derivations (the latter was not well constrained previously). In any case, it is again clear that the line of sight interstellar emission is an overwhelming source of contamination, with the searched-for SNR signal at the few percent level.  Identifying cold SNR dust will be challenging even with the improved resolution of \textit{Herschel} working close to the peak of the cold dust emission.

\subsection{Other Measures of Interstellar Column Density}\label{foreground}

A supplementary approach would be to look at other measures of the ISM
column density as a surrogate for the morphology and brightness level
expected for the contaminating interstellar dust emission.

One potential surrogate for column density is molecular line
emission. Molecular emission is affected variously by optical depth,
abundance, and excitation conditions, and so a precise one-to-one
correspondence with dust emission would not be expected.
Nevertheless, even Figure~\ref{12co}, the integrated line emission
from the FCRAO $^{12}$CO~($1-0$) data cube in the CGPS \citep{CGPS},
looks remarkably similar, though not identical, to the cold dust
emission over this extended region (Figures~\ref{maps}c--f) and to the
column density map (Figures~\ref{model_maps}b and e).
Near the SNR, the integrated CO~($2-1$) map by \cite{liszt_lucas_1999}
also shows features recognizable in the dust emission, including the
western peak and the separated south-east ridge.  \cite{Wilson_Batrla_2005} emphasize how $^{13}$CO~($1-0$) also projects on the SNR.
These tracers
suggest, as does the larger CGPS map, that the dust emission from the
interstellar cloud C should extend beyond the SNR to the west.
Note also the molecular emission toward the center of the SNR in these
maps.
The correlation of dust emission and molecular emission is not and
cannot be expected to be perfect, but it is clear that interstellar line of sight dust emission is a major contaminant.

If the interstellar material seen along the line of sight is
substantially in the foreground of the Cas~A SNR emission, as appears
to be the case, then measures of absorption can be used to trace column
density.  There are three approaches:
(i) molecular lines (OH: \citealt{Bieging_1986}; methanol:
\citealt{Reynoso_2002}) and the \ion{H}{1} 21-cm line
(e.g., \citealt{Keohane_1996}),
(ii) optical reddening \citep{Fesen_2006b}, and
(iii) X-ray absorption
\citep{Keohane_1998,Vink_1999,Hughes_2000,Willingale_2002}.
These give qualitatively the same results:
(i) the column density is highest toward the west of the SNR, where we
see the residual at 100 and 140 \micron\ after subtracting the tepid
Gaussian component (\S~\ref{internal_dust}),
(ii) there are patchy enhancements along a band extending from the
west to the south-east, thus covering the south-west and south portion
of the remnant, like feature C in our cold column density map and the
molecular emission,
(iii) the absorption is lowest across the north-east, but
(iv) there is a substantial column everywhere, including toward the
center (and the neutron star), and
(v) the surrogate column density is about the same as derived in the
above component separation, which indicates that our choice of
$\kappa_{\rm{100}}$ was reasonable.

Quantitatively, column densities from the extinction surrogates are of order $10^{22}\ {\rm cm}^{-2}$, in these units about 2.3 toward the western maximum and 1.3 toward the center ($A_V \sim 7$)\footnote{Note that these column densities are large enough to have an extinction   effect on the infrared spectra of the SNR, although no absorption   (e.g., from silicates) was detected or modeled by \cite{Rho_2008}.}. Not only are these values and the low contrast in column density comparable to what we found independently above, but there is also morphological similarity.  Compare, for example, Figure~1c in \cite{Keohane_1996} and Figure~4 in \cite{Willingale_2002} to our cold dust column density map in Figure~\ref{model_maps}e.  This leaves little room for an additional contribution at long wavelengths from the Gaussian ``tepid'' SNR dust component or a colder one.

\citet{Krause_2004_CasA} show that the patchy OH absorption correlates
well with the residual 850\,\micron\ emission seen with SCUBA (after
subtracting the synchrotron emission), and on subtracting this
foreground emission (scaled using $T_{\rm{d}} = 14$\,K) from their
160\,\micron\ map they find only a small residual and hence the
above-mentioned upper limit of 0.2\,M$_\odot$ of cold dust associated
with the SNR.

Accounting for and removal of the (foreground) emission by
correlations with extinction or molecular emission does seem a
promising technique. It will be interesting to see if it
proves useful pixel by pixel with forthcoming high signal to noise
data or only statistically.  Clearly the higher resolution imaging
anticipated with \textit{Herschel} will be essential for making
progress along these lines.

\section{Conclusions}\label{conclusions}

\begin{enumerate}
\item{We presented far-infrared/submillimeter data at 65 --
  500\,$\mu$m for the Cas~A supernova remnant and the surrounding
  region. We used these maps to characterize the interstellar dust
  emission using data from cloud regions well beyond the SNR.}
\item{We used high resolution ARAKI data to probe the spectral region between the hot dust emission from the SNR shock-front and cold dust emission. Using a spectrum-informed clean technique we identified a new tepid dust population at temperature of $\sim$35\,K. The mass of this individual dust population was estimated to be 0.06\,M$\odot$, but with considerable uncertainty because of its dependence on the choice of $\kappa$.}
\item{The dust yield for this new and independent tepid component is comparable to that estimated previously for the hot dust component by \cite{Rho_2008}. While such yields could contribute to the dust masses seen in high redshift galaxies, they are still less than the required level 0.4-1\,M\subsun estimated by \citet{Dwek_2007}. While the mass we measure is insufficient to account for the dust observed at high redshift, when taken in combination with the hot and cold dust masses previously reported by \cite{Rho_2008} and \cite{Dunne_2009}, it strengthens the argument for supernovae as a potentially significant source of dust production in the high-redshift universe.}
\item{We developed a simple physically-motivated model of the SED of the SNR and interstellar emission and fit this to six-wavelength bands at each pixel. From this we obtained temperature and dust mass column density maps.  The interstellar dust was found to be at a temperature of $\sim 16.5$\,K, in keeping with previous measurements, but now better constrained due to the improved wavelength coverage.}
\item{We show that the high level of confusion arising from the   interstellar cloud structure projected on the SNR precludes a   significant detection of cold dust directly associated with Cas~A.   The same source of confusion will have affected previous estimates   of cold dust in Cas~A, increasing the uncertainty of those   estimates.  This analysis was not sufficiently sensitive to identify the lower limiting mass found by \cite{Dunne_2009} or the lower value by \cite{Krause_2004_CasA} and therefore does not preclude the possibility of a significant population of cold SNR dust grains with temperature close to that of the interstellar dust. The higher angular resolution data anticipated with \textit{Herschel} working close to the peak of the cold dust emission, together with correlations with surrogates of the interstellar column density, could result in a more sensitive probe.}
\end{enumerate}

\acknowledgments We acknowledge the support of NASA through grant numbers NAG5-12785, NAG5-13301, and NNGO-6GI11G, the NSF Office of Polar Programs, the Canadian Space Agency, the Natural Sciences and Engineering Research Council (NSERC) of Canada, the UK Science and Technology Facilities Council (STFC), and Korea Science and Engineering Foundation (R01-2007-000-20336-0, F01-2007-000-10048-0). This work is also based on observations with AKARI, a JAXA project with the participation of ESA. Finally, we acknowledge M. Wright for the use of his 83 GHz radio map of Cas~A.

\bibliographystyle{astron}
%
%
%


\def\jnl@style{\it}
\def\aaref@jnl#1{{\jnl@style#1}}

\def\aaref@jnl#1{{\jnl@style#1}}

\def\aj{\aaref@jnl{AJ}}                   
\def\araa{\aaref@jnl{ARA\&A}}             
\def\apj{\aaref@jnl{ApJ}}                 
\def\apjl{\aaref@jnl{ApJ}}                
\def\apjs{\aaref@jnl{ApJS}}               
\def\ao{\aaref@jnl{Appl.~Opt.}}           
\def\apss{\aaref@jnl{Ap\&SS}}             
\def\aap{\aaref@jnl{A\&A}}                
\def\aapr{\aaref@jnl{A\&A~Rev.}}          
\def\aaps{\aaref@jnl{A\&AS}}              
\def\azh{\aaref@jnl{AZh}}                 
\def\baas{\aaref@jnl{BAAS}}               
\def\jrasc{\aaref@jnl{JRASC}}             
\def\memras{\aaref@jnl{MmRAS}}            
\def\mnras{\aaref@jnl{MNRAS}}             
\def\pra{\aaref@jnl{Phys.~Rev.~A}}        
\def\prb{\aaref@jnl{Phys.~Rev.~B}}        
\def\prc{\aaref@jnl{Phys.~Rev.~C}}        
\def\prd{\aaref@jnl{Phys.~Rev.~D}}        
\def\pre{\aaref@jnl{Phys.~Rev.~E}}        
\def\prl{\aaref@jnl{Phys.~Rev.~Lett.}}    
\def\pasp{\aaref@jnl{PASP}}               
\def\pasj{\aaref@jnl{PASJ}}               
\def\qjras{\aaref@jnl{QJRAS}}             
\def\skytel{\aaref@jnl{S\&T}}             
\def\solphys{\aaref@jnl{Sol.~Phys.}}      
\def\sovast{\aaref@jnl{Soviet~Ast.}}      
\def\ssr{\aaref@jnl{Space~Sci.~Rev.}}     
\def\zap{\aaref@jnl{ZAp}}                 
\def\nat{\aaref@jnl{Nature}}              
\def\iaucirc{\aaref@jnl{IAU~Circ.}}       
\def\aplett{\aaref@jnl{Astrophys.~Lett.}} 
\def\apspr{\aaref@jnl{Astrophys.~Space~Phys.~Res.}}
\def\bain{\aaref@jnl{Bull.~Astron.~Inst.~Netherlands}} 
\def\fcp{\aaref@jnl{Fund.~Cosmic~Phys.}}  
\def\gca{\aaref@jnl{Geochim.~Cosmochim.~Acta}}   
\def\grl{\aaref@jnl{Geophys.~Res.~Lett.}} 
\def\jcp{\aaref@jnl{J.~Chem.~Phys.}}      
\def\jgr{\aaref@jnl{J.~Geophys.~Res.}}    
\def\jqsrt{\aaref@jnl{J.~Quant.~Spec.~Radiat.~Transf.}}
\def\memsai{\aaref@jnl{Mem.~Soc.~Astron.~Italiana}}
\def\nphysa{\aaref@jnl{Nucl.~Phys.~A}}   
\def\physrep{\aaref@jnl{Phys.~Rep.}}   
\def\physscr{\aaref@jnl{Phys.~Scr}}   
\def\planss{\aaref@jnl{Planet.~Space~Sci.}}   
\def\procspie{\aaref@jnl{Proc.~SPIE}}   

\let\astap=\aap
\let\apjlett=\apjl
\let\apjsupp=\apjs
\let\applopt=\ao

\bibliography{references}

\end{document}